# HIGH REVERSE BREAKDOWN a-C:H/Si DIODES MANUFACTURED BY rf-PECVD

Shashi Paul[*] and F.J Clough
Emerging Technologies Research Centre
De Montfort University, Hawthorn Building, The Gateway, Leicester LE1 9BH, UK

**ABSTRACT**

Thin films of hydrogenated amorphous carbon ( a-C:H ) deposited by radio frequency plasma-enhanced chemical vapour deposition ( rf-PECVD ) have been studied for various applications. An interesting property of these films is their high breakdown strength ( $10^7$ Vcm$^{-1}$ ). This property of a-C:H can be exploited in high breakdown heterostructure diodes or as passivation layers and insulator layers in MIS devices. Reports on the applications of a-C:H/Si diodes exist in the literature. Diodes in which the a-C:H films have been deposited by rf-PECVD, have been reported only once. In this article the diodes produced reportedly failed to exhibit reproducible I-V characteristics under high voltage stress. We have investigated the process dependence of structural and electrical properties of rf-PECVD a-C:H films deposited at room temperature from a $CH_4$/Ar gas mixture (at a pressure of 100 mTorr) using a capacitively coupled rf-PECVD. We observe a clear correlation between the dc-self bias and the rectification ratio of a-C:H/Si heterojunction diodes. Optimised diodes show rectification ratios upto $10^4$ and a stable reverse breakdown voltage, typically around 850 V. I-V and C-V measurements show no evidence of hystersis. Scanning Electron Microscopy was carried out to determine the quality of the films deposited. Micro-Raman analysis was used to estimate the $I_D/I_G$ ratio in the films deposited under different dc-self bias.

**INTRODUCTION**

The applications of hydrogenated amorphous carbon (a-C:H) films to electronic devices have been of considerable interest in recent years [1-4]. An interesting property of these films is their high breakdown strength ($10^7$ Vcm$^{-1}$). Properties such as a low dielectric constant (which implies higher switching speed), high breakdown strength and deposition at room temperature, make this material useful for cheap, glass compatible technologies and for power electronics. One very widely used class of device that could benefit from these properties of is the p-n diode. The electrical characteristics of p-n junctions are fundamental to the electronics industry. Rectification, isolation, amplification and switching are key operations that involve the use of p-n junctions in electronic circuits. Therefore, fabrication of p-n junction diodes is an appropriate starting point in the development of a semiconductor technology based on amorphous carbon. In order to implement a-C:H films (deposited by rf-PECVD) in heterostructure junctions, it is vital to understand the electrical behaviour of a p-n heterojunction as a function of the growth parameters. The plasma deposition of a-C:H proceeds by ion bombardment which is activated by the negative dc self-bias ($V_{SB}$) of the driven electrode on which the substrate is placed [5]. The film properties are therefore strongly influenced by the magnitude of the dc self-bias which is in turn dependant on the applied rf power and the operating pressure. The effects of dc self-bias on resistivity, activation energy, optical bandgap, hydrogen content and spin density have been demonstrated by Ohta et al [6]. Therefore, understanding the a-C:H/Si junction as a function dc-self bias will shed light on its electrical behaviour. This understanding can then be used to realise high reverse breakdown diodes.

---

[*] Present address: Department of Engineering, Cambridge University, Cambridge CB2 1PZ, UK

The present paper investigates the rectification ratio of a-C:H/Si hereostructre as a function of dc self-bias and uses this knowledge to fabricate diodes with high reverse breakdown voltage.

**EXPERIMENTAL**

Thin films of a-C:H were deposited at room temperature from $CH_4$/Ar gas mixture using a capacitively coupled rf-PECVD (13.56 MHz) system at a pressure of 100 mTorr. The plasma power was varied to develop dc self-bias voltages ranging from -45 V to -240 V. A nominal a-C:H thickness of 90 nm was maintained for all samples to investigate the rectification as a function dc-self bias voltage. The heterojunction diodes were manufactured by depositing a-C:H on to n-type and p-type Si substrates, of resisivities 1-2 Ωcm and 200-220 Ωcm respectively. The upper contact was formed by evaporating Al through a shadow mask with a circular aperture of diameter 1 mm. The bottom contact was formed by blanket evaporation of an Al layer on to the back of the Si substrate followed by annealing at 450$^o$C for 30 minutes prior to the deposition of a-C:H. This was done to ensure the ohmic nature of the bottom contact. Current-voltage (I-V) characteristics for the a-C:H/Si structures were measured using a pc-driven picoammeter (HP4140B) and a high voltage source (HP4284A). During the measurements of the I-V characteristics, the dc voltage was applied to the upper contact while the back contact was earthed. To realise the high reverse breakdown voltage, a-C:H/Si diodes were prepared with two different thicknesses, 90 nm and 200 nm. Scanning Electron Microscopy (SEM) was used to determine the surface quality of the deposited films. Micro-Raman measurements were carried out using the 514.5 nm line of an Ar ion laser with a 2 mm spot size.

The heterojunction diodes were also exposed to prolonged electrical stress (upto 2 hours) to investigate charge trapping either at the Al/a-C:H interface or in the bulk. Significant charge trapping is expected to result in poor reproducibility in the I-V characteristics [7].

**RESULTS AND DISCUSSION**

The I-V behaviour of a-C:H/Si heterostructures manufactured on to 1-2 ohm-cm n-type substrates, at different dc-biases, is shown in Fig.-1. A typical symmetrical characteristic of a device manufactured by depositing at $V_{SB}$ = -70 V is shown in Fig.1(a). The rectification ratio ( R.R) measured at 10 V for these devices is 1. An increase in the dc self bias to -95 V shows slight rectification ( Fig.-1(b)). The highest rectification ratio of 105 is achieved at $V_{SB}$ = -200 V in this investigation. To understand the effect of $V_{SB}$ on device behaviour we plotted the rectification ratio and resistivity of the films as functions of $V_{SB}$ in Fig.-2. The resistivity of the films falls from $2 \times 10^{12}$ to $10^6$ Ωcm as the dc self-bias is varied from –70 V to -200 V.

The decrease in the resistivity can be explained on the basis of structural changes arising from a change in the dc self-bias. This was confirmed by Raman-spectroscopy and optical band gap analysis. The explanation of symmetrical I-V behaviour at lower dc-self bias is given by the fact that the resistivity of the films is very high ($10^{12}$ Ωcm) and the devices behave like MIS structures. As the resistivity starts falling and the semiconducting regime is approached (resistivity < $10^7$ Ωcm) the increase in the rectification behaviour is obvious. This novel investigation shows how a single parameter, namely the dc self-bias, changes the device properties strongly (from the MIS to the heterostructure configuration). On the basis of this investigation, we selected the growth conditions which corresponded to the highest rectification ratio. This was done to realise the high reverse breakdown diode.

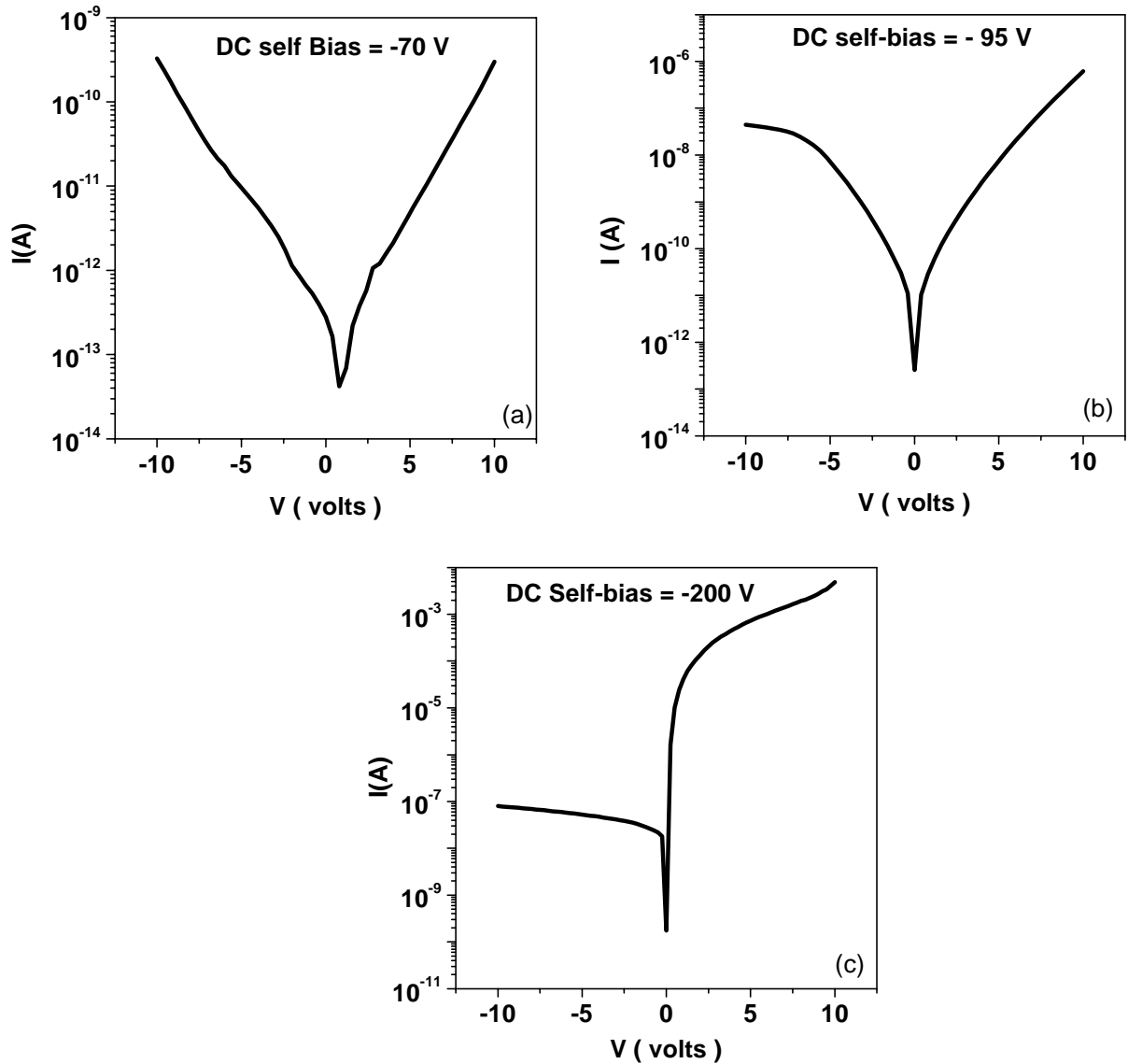

Fig.- 1 Variation in I-V behaviour of a-C:H/Si heterostrucutres diodes as a function of dc self-bias

The I-V behaviour of an a-C:H/Si heterostrucutre is shown in Fig-3. The asymmetrical (at positive and negative applied bias) I-V curves are clearly indicative of diode behaviour. These devices were manufactured to obtain high reverse breakdown voltages by depositing a-C:H on to 200-220 Ωcm p-type Si substrates. Fig.-3(a) shows the typical I-V behaviour of diodes produced by depositing a 90 nm thick film of a-C:H. The reverse breakdown voltage and rectification ratio of these diodes were 400 V and $10^4$ (measured at 50 V) respectively. The reproducibility of these characteristics was studied by keeping these devices continually under increasing electrical stress (from - 200 V to 350 V for 2 hours). From Fig.-3(a) we can see that there is no change in the electrical behaviour of these devices arising from the application of this continuous electrical stress.

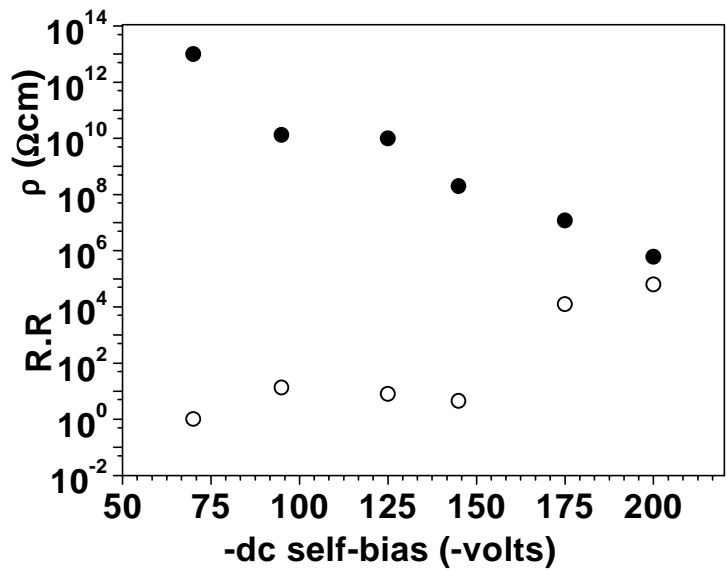

Fig.- 2 Variation in resistivity (measured at 1MVcm$^{-1}$) and rectification ratio (at 10 V) with dc-self bias; (•) filled circles represented variation of resistivity and unfilled circles (o) depict the change in the rectification ratio with the dc self bias.

Diodes of higher reverse breakdown strengths were achieved by fabricating heterostructures with a 200 nm a-C:H layer. Typical I-V behaviour of such diodes is as shown in Fig.3(b). The reverse breakdown voltage of such diodes was typically 850 V. The electrical characteristics of such diodes were again unaltered following the application of prolonged electrical stress (Fig.-3(b)).

An earlier attempt to investigate the high-voltage performance of a-C:H/Si diodes by Amaratunga *et al* [7] resulted in poor reproducibility after the application of high voltage. In this report it was suggested that this was due to an excessive increase in the barrier height as a result of fixed negative charges at the Al/a-C:H interface and in the bulk of a-C:H. Trapped charges would be indicated by hystersis in the I-V or C-V characteristics of the devices that we studied. If the explanation given in [7] is true then we can say that the trapped charge densities in our devices, both at the interface and in the bulk, are negligibly small and have no significant influence on device behaviour. Production of a-C:H/Si diodes (for high voltage operation) using the technique of photochemical vapour deposition has also been reported [8]. This technique does not expose the growing film to ion bombardment and, it is suggested, results in reduced trapped charge densities. The authors of [8] have successfully demonstrated that high voltage diodes prepared by this technique can withstand reverse bias voltages upto 400 V.

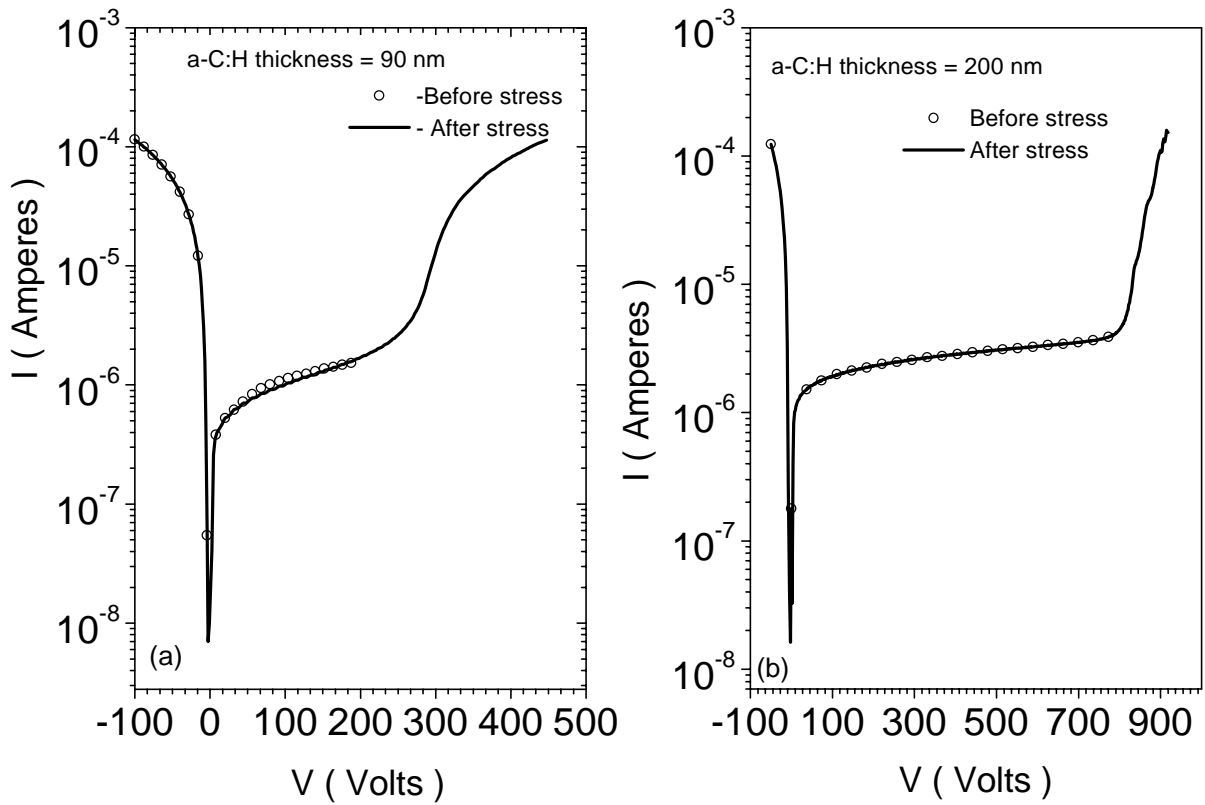

Fig.-3 The circles show the I-V behaviour measured before the application of electrical stress, while the line shows after the application.

Through optimisation of the growth conditions using detailed investigation of the film properties a more robust diode has been realised by rf-PECVD. In their report [7], a number of large crystallites of diamond, of size between 100 nm and 200 nm, were observed in the amorphous carbon matrix. The crystallites in the material will lead to preferred conduction paths through the amorphous carbon layer or less resistive paths around the diamond crystallites and, ultimately, premature breakdown. To check for large crystallites and pin-holes in our films we extensively investigated the a-C:H films used in these devices using SEM. Fig-4 shows the homogenous and smooth surface of the a-C:H films deposited in our laboratory. The films were uniform and free from pinholes. Therefore, to achieve high breakdown voltage a-C:H/Si diodes the important factors appear to be uniformity of the a-C:H films and low trap states. These can be achieved by detailed optimisation of the PECVD growth conditions.

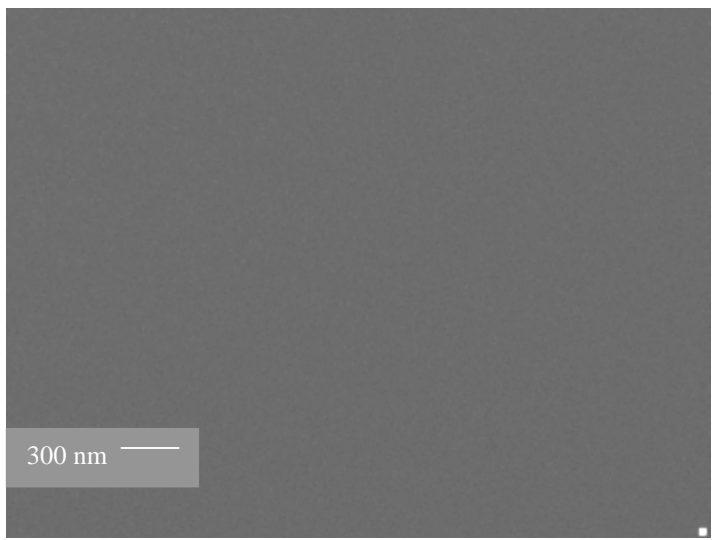

Fig.-4 SEM micrograph of an a-C:H film deposited on a Si (100) substrate.

**SUMMARY**

In summary, a high reverse breakdown voltage a-C:H/Si diode has been successfully fabricated by rf-PECVD for the first time. The trap states (both at the interface and in the bulk) can be reduced by selecting the growth conditions carefully. Optimum films were free from any inclusion of diamond crystallites and pinholes.